# High Pressure Growth of Transition-Metal Monosilicide RhGe Single Crystals

Xiangjiang Dong(董祥江)[1,†], Bowen Zhang(张博文)[1,†], Xubin Ye(叶旭斌)[2], Peng Wei(魏鹏)[1], Lie Lian(廉磊)[1], Ning Sun(孙宁)[1], Youwen Long(龙有文)[2], Shangjie Tian(田尚杰)[3], Shouguo Wang(王守国)[3], Hechang Lei(雷和畅)[4,5,*], Runze Yu(于润泽)[1,*]

[1]Center for High-Pressure Science and Technology Advanced Research, Beijing 100093, China

[2]Beijing National Laboratory for Condensed Matter Physics, Institute of Physics, Chinese Academy of Sciences, Beijing 100190, China

[3]Anhui Provincial Key Laboratory of Magnetic Functional Materials and Devices, School of Materials Science and Engineering, Anhui University, Hefei 230601, China

[4]School of Physics and Beijing Key Laboratory of Optoelectronic Functional Materials & MicroNano Devices, Renmin University of China, Beijing 100872, China

[5]Key Laboratory of Quantum State Construction and Manipulation (Ministry of Education), Renmin University of China, Beijing 100872, China

†These authors contributed equally to this work.

*Corresponding author: hlei@ruc.edu.cn; runze.yu@hpstar.ac.cn

**Abstract:**

Transition-metal monosilicide RhGe has been reported to exhibit weak itinerant ferromagnetism, superconductivity, and topological properties. In this study, we report the high-pressure growth of high-quality RhGe single crystals up to millimeter size using flux method. Transport measurements reveal the metallic behavior of RhGe between 2–300 K with Fermi liquid behavior at low temperature region. However, no superconductivity was observed with variations in Ge composition. Magnetic characterizations indicate that RhGe exhibits a paramagnetic behavior between 2–300 K. The high-quality, large-size RhGe single crystals pave the way for further investigation of their topological properties using spectroscopic techniques.

PACS：07.35.+k;81.10.-h;

**Introduction:**

Transition metals monosilicides MX (M = Cr, Mn, Fe, Co, Ru, Rh, and Os; X = Si, Ge) crystallized in FeSi-type structure (B20 structure, $P2_13$), featuring a cubic centrosymmetric lattice, have attracted significant attention recently to their intriguing magnetic, topological, and transport properties [1-6]. For instance, MnSi exhibits an itinerant helical magnetic structure below $T_C$ = 30 K, transitioning to a conical structure under a magnetic field and eventually evolving into ferromagnetism [7-10]. Additionally, a skyrmion phase has been observed in the conical phase near $T_C$ [11-12]. This magnetic structure can be suppressed under pressure, with a quantum critical point observed at approximately 1.46 GPa [8-9]. Another notable compound CoSi demonstrates nontrivial electronic structure topology [13-15]. The electronic structure with nonzero topological charge and Fermi arcs, connecting projections of band-touching nodes at Γ and R points on the surface Brillouin zone, has been observed in angle-resolved photoemission spectroscopy (ARPES) experiments [16-18]. Besides fascinating magnetic and topological properties, transition metals monosilicides also exhibit excellent thermoelectric performance. For example, CoSi shows Seebeck coefficient at room temperature of about -80 μV/K at stoichiometric composition [19]. Power factor up to 5.5 mW/mK$^2$ has been observed in CoSi at 350–400 K, surpassing that of the well-known thermoelectric material $Bi_2Te_3$ [19].

Despite the many exotic physical properties discovered in B20-structured materials, superconductivity remains rare. Tsvyashchenk *et al.* reported that RhGe showed a superconducting state below 4.3 K and weak itinerant-electron ferromagnetism below 140 K [20]. However, no further investigations have been conducted on this compound since then. The absence of inversion symmetry in the superconducting order parameter results in a mixture of spin-singlet and spin-triplet components [21], making RhGe possibly be a novel platform for studying *p*-wave superconductivity. Moreover, the potential coexistence of itinerant ferromagnetism and superconductivity in RhGe provides a unique opportunity to explore the intrinsic physical properties of unconventional superconductivity [20]. Recent ARPES experiments have observed band-touching nodes in the bulk electronic spectrum and Fermi arcs in CoSi [16-18]. However, the small spin-orbit coupling (SOC) in CoSi makes it challenging to resolve closely lying pairs of Fermi arcs experimentally, and the energy positions of these nodes and the peculiarities of the corresponding surface Fermi arcs must be determined through band structure calculations [16-18]. Given that the SOC of RhGe is significantly larger than that of CoSi, with node splitting nearly twice as large [2], RhGe also offers an opportunity to study the evolution of band structures and Fermi arcs with the enhancement of SOC.

To address these issues, high-quality RhGe single crystals are essential. High-quality single crystals are indispensable for determining electronic structures using ARPES, and they often reveal intrinsic properties that cannot be observed in polycrystalline samples. In this work, we report the growth of RhGe single crystals under high pressure and high temperature, along with detailed characterization of their physical properties.

**Experimental details:**

Singel crystals of RhGe were grown using high pressure cubic anvil apparatus. The precursors of Rh (99.99%) and Ge (99.99%) with mole ratio 1:2 was thoroughly ground and mixed. Additional Ge was added as a flux due to its low melting point. The precursor was sealed into an hBN capsule (Au and Pt capsules were avoided as they can react with Ge under high pressure and high temperature), then placed into the center of the high-pressure cells and compressed up to 5 GPa. The temperature was first rapidly increased to 1200 °C with rate 6 °C /min and maintained at that temperature for 4h followed by a gradual decrease to 600 °C over 30 hours.

Finally, the sample was quenched to room temperature before releasing the pressure. The single-crystal quality and orientation were verified using single-crystal X-ray diffraction on a high-resolution system (SmartLab) with Cu $K_\alpha$ radiation ($\lambda$ = 1.5406 Å), and the Laue pattern was collected in backscattering mode. Powder X-ray diffraction was conducted at room temperature, using an X-ray diffractometer (D/max 2500V) with Cu $K_\alpha$ radiation at operating voltage and current of about 40 kV and 150 mA, respectively. The structure was refined using the Rietveld method via the GSAS program [22]. The magnetic properties were measured using the MPMS3 of Quantum Design under magnetic field of 1 T in zero field cooling and field cooling modes. Scanning Electron Microscope (SEM) was carried out by JEOL, JSM-7000F. Chemical compositions of single crystals were determined using a scanning electron microscope with an energy dispersive X-ray analyzer. Transport measurement was performed using four-probe method. Specific heat was measured using a PPMS.

**Results and Discussion:**

Given the low melting point of Ge element, we chose Ge as the flux to growth RhGe single crystal. Initial attempts with a Rh : Ge ratio of 1 : 1 yielded single crystals up to 1.0×0.5×0.5 mm$^3$ in size, but the quality was suboptimal. By adjusting the initial ratio to 1 : 2 to increase the content of flux, we successfully obtained high-quality RhGe single crystals. Figure 1(a) shows a photograph of single crystals of RhGe up to millimeter size. It can be seen the crystals show dark color with shining surface (See Figure 1(a)). X-ray powder diffraction can be fit very well using the $P2_13$ structure (See Figure 1b). Energy-dispersive X-ray (EDX) analysis confirmed the composition of Rh and Ge to be 1 : 1, which is highly coincided with that nominal composition (See Figure 1(c)). Figure 1d shows a Laue back reflection pattern on RhGe single crystal for a high symmetrical plane of (100). Sharp diffraction spots were observed without visible spallation, and all the spots were well-indexed based on a cubic symmetry, suggesting the high quality of RhGe single crystals.

The resistivity of RhGe single crystal $\rho(T)$ exhibits metallic behavior between 25 - 300 K, with gradual upturn below 25 K, indicating the possibility of weak localization. There was no observation of drop indicating the appearance of superconductivity. The low temperature range of $\rho(T)$ curve between 25 - 70 K follows the $T^2$ dependence, consistent with Fermi liquid behavior, rather than the $T^3$ behavior reported in previous studies. As no superconductivity was observed down to 2 K, we believed the absence of superconductivity may be intrinsic and not due to Ge content variations as previously reported.

Next, we focus on the magnetism of RhGe single crystals. Temperature dependent of magnetic susceptibility $\chi(T)$ at magnetic field of 1 T indicates that RhGe single crystal shows a paramagnetism in the whole temperature range (See Figure 3(a)), and the ferromagnetism with $T_C \sim 140$ K [20] is absent. Moreover, field-dependent magnetization $M(\mu_0H)$ curve at 5 K shows a nearly liner behavior (See Figure 3(b)), indicating no magnetic order at low temperature, consistent with paramagnetic behavior shown in $\chi(T)$ curve.

Specific heat measurement of RhGe single crystal is shown in Figure 4. No significant jump either at 4.3 K or 140 K can be observed. This result further confirms the absence of superconductivity and ferromagnetism for RhGe single crystals.

Since no superconductivity was observed in pristine RhGe, carrier doping or isovalent substitution may be a possible way to realize superconductivity in this system. Previous study shows that the orthorhombic structure IrGe exhibit superconductivity with $T_c \sim 4.7$ K [23]. It would be interesting to study whether the Ir doping in RhGe can induce superconductivity. Besides superconductivity, the larger SOC of Ir may also have some influence on the surface state and lead to novel topological properties. These studies will provide further opportunities to gain deeper insights into the intrinsic properties of B20-structured materials.

**Summary:**

We successfully growth the high-quality RhGe single crystals up to millimeter size under high pressure and high temperature. Transport measurement reveals the metallic behavior and Fermi liquid behavior of RhGe single crystal at low temperature, while magnetic measurements indicate that RhGe is a paramagnet. Neither superconductivity nor ferromagnetism is observed down to 2 K. This high-quality RhGe single crystals provide a promising platform for further investigation of its topological properties using APRES. Moreover, this study highlights the ongoing challenge of discovering superconductivity in B20-structured compounds.

**Acknowledgment:**

This work was supported by National Key Research and Development Program of China (Grant Nos. 2023YFA1406000, 2022YFA1403800, 2021YFA1400300 and 2023YFA1406500), the National Natural Science Foundation of China (Grant Nos. 12474002, 22171283, 12425403, 12261131499, 12304268 and 12274459), China Postdoctoral Science Foundation (Grant Nos. 2023M730011, 2023M743741).

**Figure captions:**

Figure 1: (a) Photograph of RhGe single crystal. (b) Rietveld refinement of the X-ray diffraction pattern of RhGe. The inset is the crystal structure of RhGe. (c) EDX spectrum for RhGe single crystal. The inset shows the SEM photo of this sample. (d) Laue diffraction spots of (100) plane for RhGe single crystal.

Figure 2: Temperature dependence of resistivity for RhGe single crystal. The red line shows the fit using the formula $\rho = \rho_0 + AT^2$.

Figure 3: (a) Temperature dependence of magnetic susceptibility $\chi(T)$ for RhGe single crystal under magnetic field of 1 T in zero field cooling and field cooling modes. (b) Field-dependent magnetization $M(\mu_0 H)$ of RhGe single crystal at 5 K.

Figure 4: Temperature dependence of specific heat for RhGe single crystal.

**Figure 1**

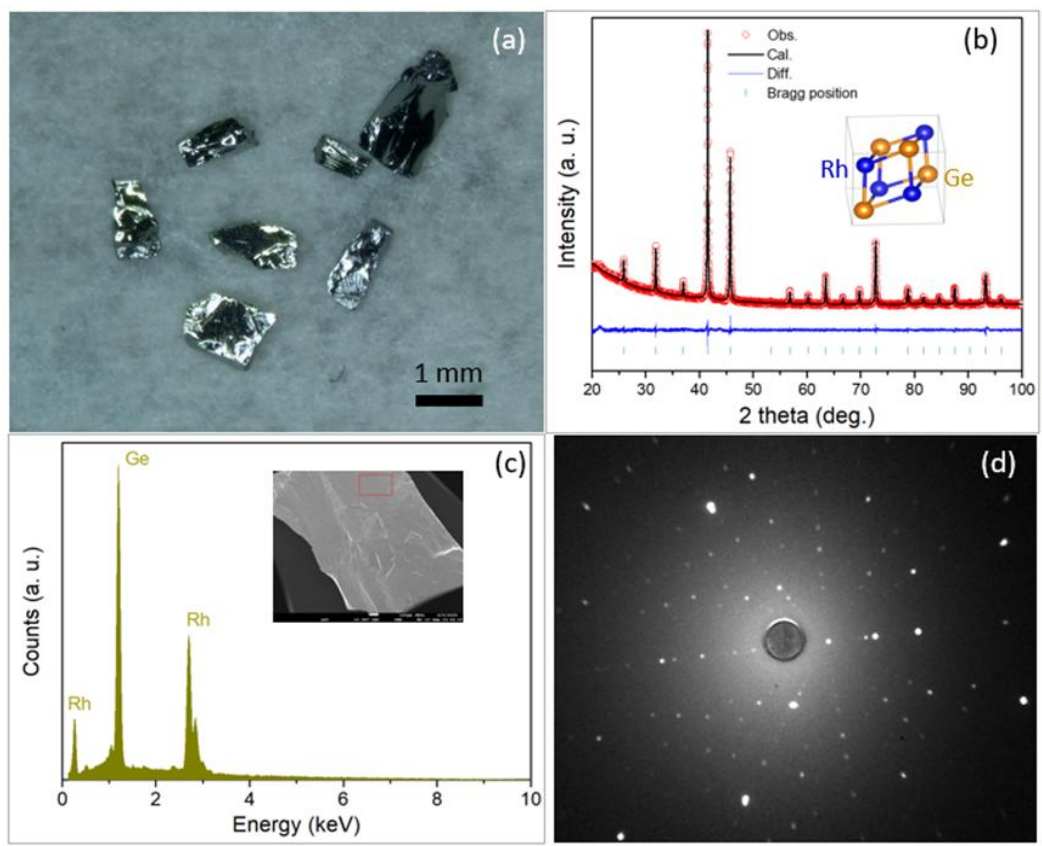

**Figure 2**

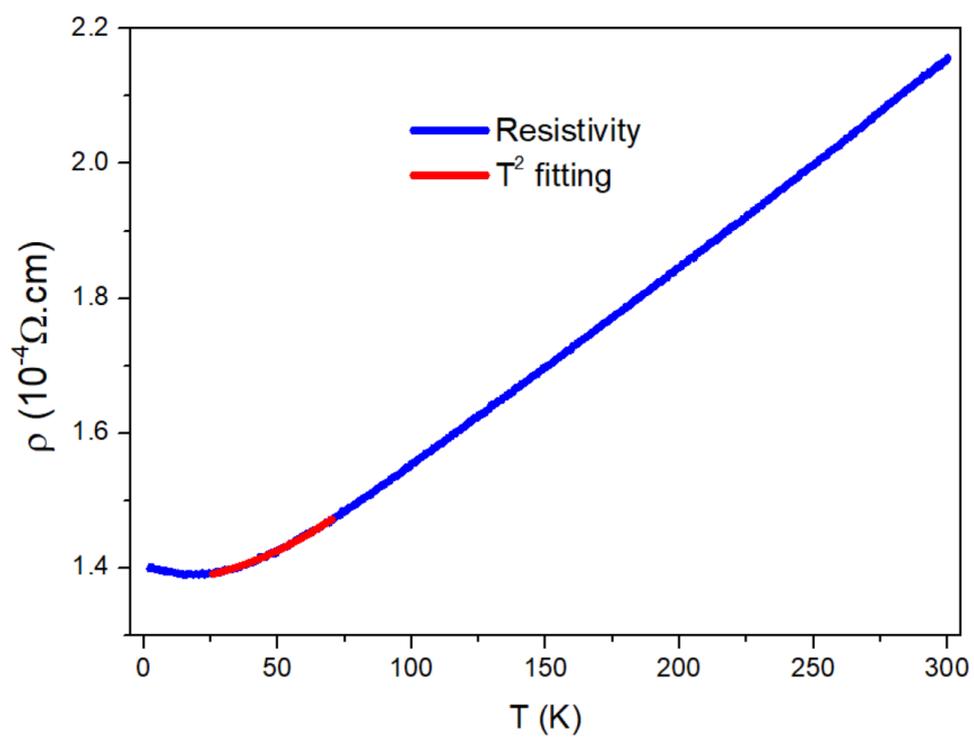

**Figure 3**

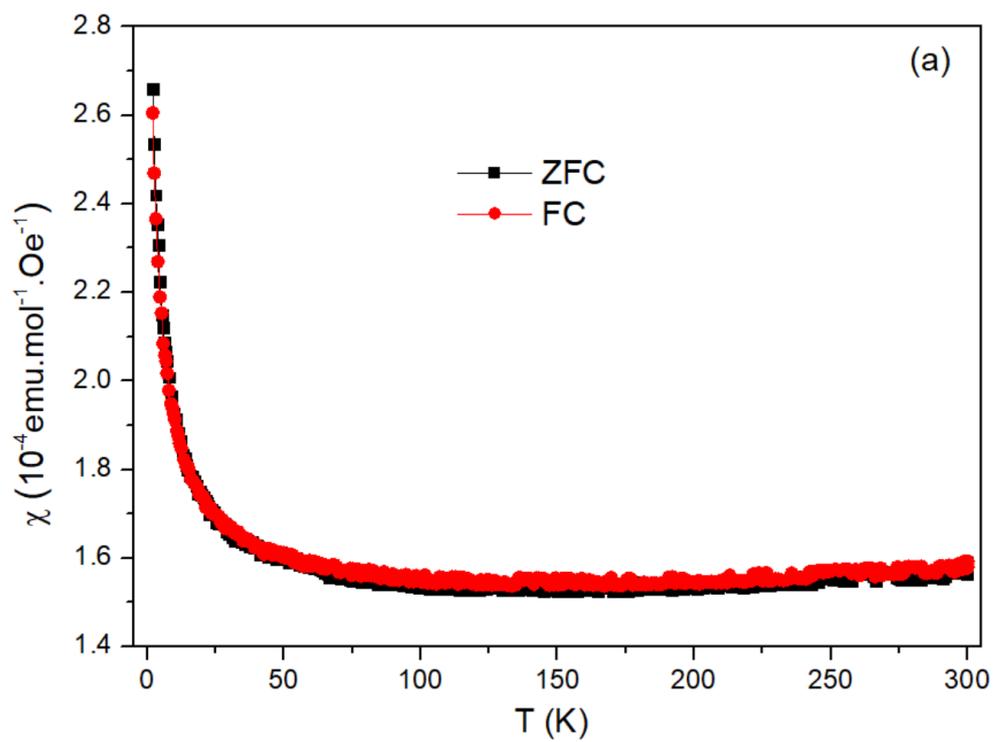

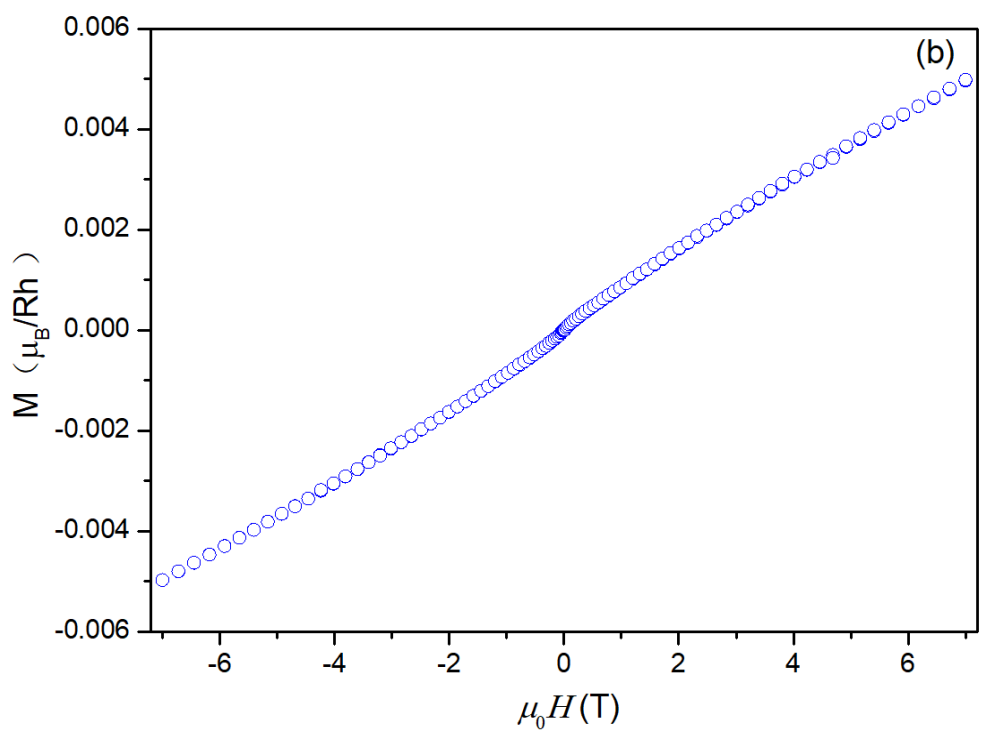

**Figure 4**

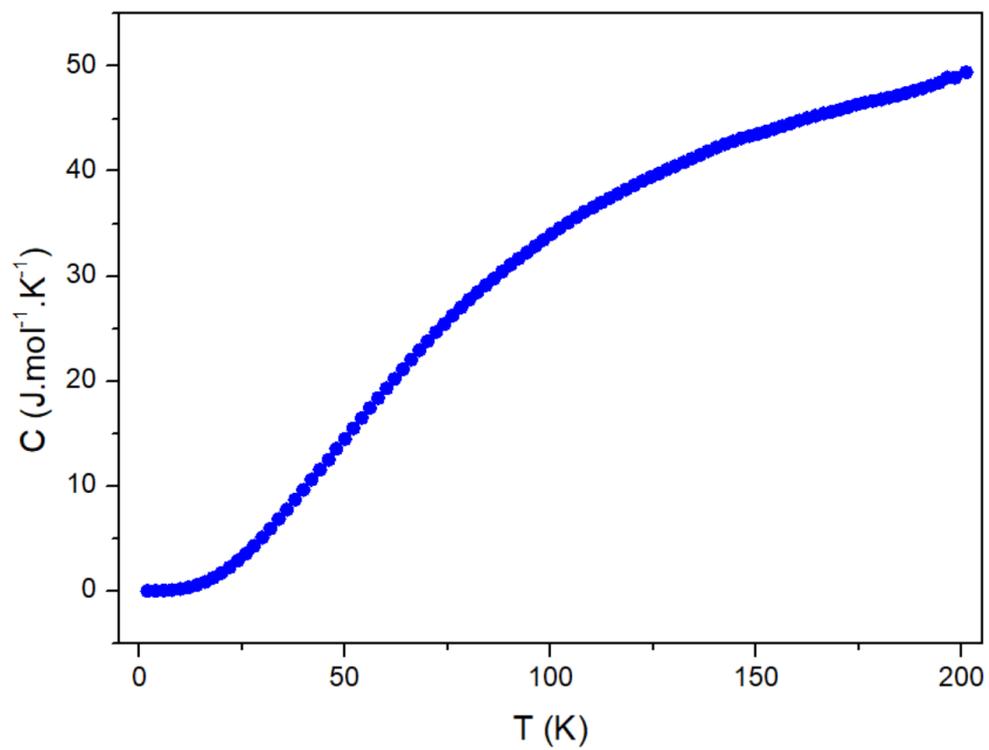